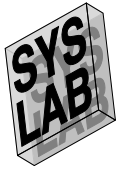 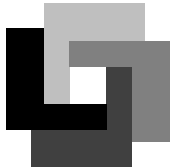

# Refining Business Processes[*]

Bernhard Rumpe, Veronika Thurner
Department of Computer Science, Technical University of Munich
Arcisstr. 21, 80290 Munich, Germany
email: {rumpe | thurner}@in.tum.de

**Abstract**

In this paper, we present a calculus for refinement of business process models, based on a precise definition of business processes and process nets. Business process models are a vital concept for communicating with experts of the application domain. Depending on the roles and responsibilities of the application domain experts involved, process models are discussed on different levels of abstraction. These may range from detailed regulations for process execution to the interrelation of basic core processes on a strategic level. To ensure consistency and to allow for a flexible integration of process information on different levels of abstraction, we introduce refinement rules that allow the incremental addition to and refinement of the information in a process model, while maintaining the validity of more abstract high level processes. In particular, we allow the decomposition of single processes and logical data channels, as well as the extension of the interface and channel structure to information that is newly gained or increased in relevance during the modeling process.

## 1 Motivation

Business process modeling today becomes an increasingly important technique for capturing system behavior on different levels of abstraction. A major strength of business process models is their suitability for supporting communication between system users and system developers, as they capture rather intuitively what the user and the system do to achieve certain goals [Öst95].

In addition to an appropriate notation, techniques are needed which effectively employ the notation in system modeling. Ideally, these techniques are supported by powerful tools that not only provide suitable editors for business process models, but also support sophisticated functionality such as the checking of appropriate context conditions, simulation or execution of business processes, or code generation [Fab98, Sch92]. Just as not every collection of words builds a meaningful sentence, not every combination of business processes is a meaningful business process net. Context conditions on business processes restrict possible combinations in such a way that the valid results are meaningful.

[*]This work originates from the SysLab project, supported by the DFG under the Leibniz program and by Siemens-Nixdorf, as well as from the FORSOFT project, supported by the Bayerische Forschungsstiftung.

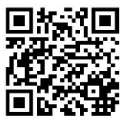

Typically in systems modeling, business process models are not developed in isolation, but in combination with other system aspects and corresponding notations, such as object models and class diagrams to describe different viewpoints on a system in an integrated way [KR94, Gro97, RBP$^+$91, Boo94]. Consequently, a smooth relation between the different concepts and their corresponding notations must be provided, to allow separate views to define an integrated, consistent system description. Today, such integrations are still under investigation, even for widespread sets of modeling notations such as the UML (for example, see [EFLR98, BGH$^+$97, KR98, BCMR98]). Thus, it is not surprising that a combination of business process models with other modeling concepts (e.g. as provided by UML) which is smoothly integrated on a formal basis and well accepted in practice does not yet exist.

One possible way of integrating notations for different modeling concepts formalizes these notations based on some well known formalism. Subsequently, the inter-relations of these notations are specified on the semantic level in terms of context conditions which ensure that the separate parts of the model fit together consistently. Furthermore, transformation rules are provided, which allow the translation of information from one modeling notation into another. For example, business process models could be transformed into corresponding interaction diagrams, state machines, or code. Of course, these transformation rules that are provided must be sound with respects to the given semantics.

When modeling real world systems, even model diagrams that focus on just one system aspect, such as data or behavior, tend to get very large, complex and thus difficult to handle. A common approach to reduce this complexity is the decomposition of the model into separate parts. Usually, these parts will be interrelated. Therefore, analogous to the transformation rules mentioned above, which relate different modeling notations, techniques are needed which relate different models within a single notation.

In this paper, we will discuss transformation rules within a notation for business process models, focussing on *refinement* as a special and widely used type of transformation within a single modeling technique. Refinement denotes the addition of more detailed information, possibly on a more detailed level, while preserving the original information. Here, preservation of some given information means that it is possibly strengthened, but never violated. A complete formal treatment of business processes, their refinement and a proof of their semantic correctness is beyond the scope of this paper. Methodical guidelines for the use of this calculus are currently under development. Our formal notion of refinement is more restrictive than some of the similar concepts used elsewhere. Refinement preserves and strengthens given properties, but does not violate them. This explicitly excludes changes at lower levels, where higher level properties are violated.

Usually, a business process model is not a big bang invention, but a step by step development. Refinement rules exhibit their power through the possibility of combining them to support more complex developments.

Section 2 introduces the concept of business process models by a small example. Section 3 presents our notion of refinement and demonstrates the refinement rules with the help of an example. Finally, Section 4 summarizes our results and presents an outlook.

## 2 A Notion of Business Processes

In this section, we will introduce briefly the concept of business processes. After informally introducing the core aspects with an example, we will give a precise definition of business processes by providing an abstract syntax.

As we focus here on refinement rules rather than on complex process structures, we restrict ourselves to exemplary system behavior throughout this work, instead of modeling alternative or cyclic processes. A treatment of complex business process structures is provided by [Thu98].

### 2.1 Introduction to Business Processes

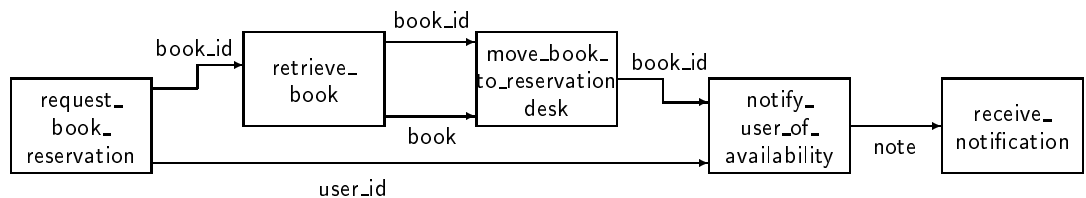

Figure 1: An example process net

We introduce an intuitive understanding of business process models by discussing the *business process net* in Figure 1. The example visualizes that part of a business process which takes place in a library when a library user requests the reservation of a book. Starting from the identifier of the requested book, the book is retrieved by a library service and forwarded to the reservation desk. Then, a notification of the availability of the requested book is issued and forwarded to the library user, who receives this notification.

The diagram depicted in Figure 1 contains a set of boxes, each of which represents a *business process* (or in short *process*). Each of these processes serves a special duty or contributes to the performance of a certain task. At an early stage of system modeling where business process modeling is often employed as a means of requirements elicitation, we are not interested in *who* performs a process, as this already includes some design decisions. As modeling and design procedes, processes are assigned roles according to pragmatic aspects such as required skill, authority or accountability. Via the concept of roles, processes are then assigned to people and other actors in the system's organizational structure. As business processes only describe one view of a system or organization, we cannot expect everything to be described with them. For example, pre- and postconditions and global state changes are not core concepts of business processes, but can be handeled by appropriate extensions.

In the early stages of process modeling, we focus on the procedural aspects of business process, i.e. on *what* has to be done to achieve some system goal, thus modeling business processes and their *causal dependencies* that are due to the exchange of information or material. Causal dependencies are represented by arrows. They may be annotated with message types, and channel names which are composed of port identifiers. Methodically it is important to allow causality constraints to be explicitly defined, as this also allows to compare these with the causalities that may be derived from invariants or code. Message types refer to business objects as defined in a data dictionary, or to data types specified in the corresponding data model and primarily denote pieces of information (or even material) flowing between the business processes. In the following, we use the terms data, pieces of information, and messages interchangeably.

Within a process net, business processes are uniquely identified by their name. As a modeling convention, the process name should provide a first intuitive notion of what the process does and ideally also its most important object.

A process communicates with other processes via its typed interface. From the perspective of methodology, when developing a business process model, at first it often is both practical and sufficient to introduce a new business process by providing its name and an informal textual explanation of its transition relation, without precisely specifiying the process interface or the processing mechanism.

The processes in a process net have a causal connection to describe which process depends on which. Such a causal dependency is indicated by an arrow. As circular dependencies must be avoided, the process net needs to be acyclic. However, each process is allowed to have several incoming and outgoing dependencies, even several between the same two processes. To ease the reading of a business process net, we recommend to direct all arrows from left to right. The *interface* of a business process is determined by the set of dependencies it is connected with. We distinguish incoming and outgoing dependencies. As with processes, dependencies may be attached with a name (type) to indicate the kind of dependency. To uniquely identify dependencies, additional identifier names may be used. In Figure 1, dependencies and their connections are depicted by arrows only. Internally, additional port names are used to define the connection structure, but these are normally omitted in a graphic representation. It usually suffices to qualify a dependency by naming source and destination process without any more information. Some of the arrows are not completely connected to processes, but indicate a dependency with the environment (either incoming or outgoing).

The business process net in Figure 1 is composed of *black-box descriptions* of several business processes. These processes can be described in more detail by regarding their *glass-box definition*, which again is a business process net that realizes at least the external interface given in the black-box description of its parent process. For example, the black-box description of a process already introduced in Figure 1 is shown in Figure 2. A corresponding glass-box definition is presented in Figure 3.

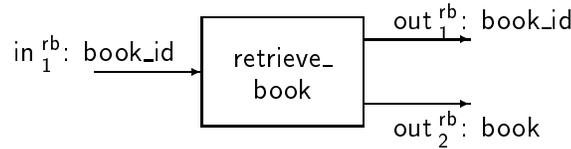

Figure 2: Black-box process description

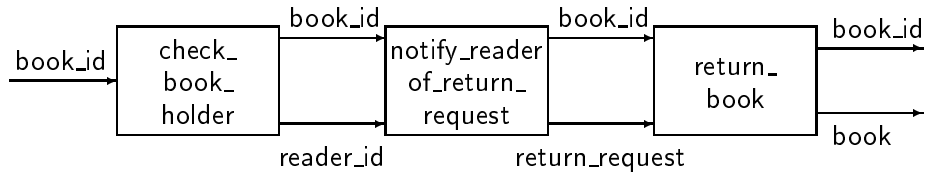

Figure 3: Glass-box process definition by a process net

This technique provides a very simple composition operation on business processes that can be used to hierarchically structure business process models into nets of different layers. It does not only allow to compose business process nets, but also to decompose a given single business process using a more detailed business process net.

Usually, causal dependencies are due to data exchange, where some data that is produced by the former process has to be used by the later process. Some rare other cases stem from some kind of synchronisation constraints, in which the former process releases a resource and the later process aquires it. This can also be modeled by data flow. Therefore we can simply interpret all dependencies as *dataflow channels* where exactly one piece of (perhaps complex) data is flowing. Then the dependency types may be interpreted as data types for the channels. In the context of dataflow diagrams, we speak of *input and output ports* which are connected by channels. Therefore, the interface of a business process is defined by a set of input and output ports.

At the beginning of the modeling process, quite often it might not be obvious which type of data should be associated with a certain dependency (dataflow channel). Consequently, the modeler is not forced to attach a type of data to each port. This would prohibit the adding of underspecified process types to the model. Instead, it is allowed to add newly gained information on the dependency to the model when appropriate. Furthermore, when modeling on a high level of abstraction, it is usually not sensible to attach data sorts to ports, as communication on this level tends to be very complex and is therefore often only vaguely understood.

In contrast to typical dataflow nets, a business process net relies on a more involved computation idea. Based on two basic assumptions, we get a computation model that reflects reality in a better way, and furthermore allows for an easy decomposition of business processes. We assume

1. Although on each channel, conceptually only one piece of data arrives, we assume that this data can be structured in a complex way and can itself be provided piecewise.

2. In general, business processes are *greedy*, i.e. trying to compute as much as possible from the partial input data that is already given. Although this is not always appropriate and typically not the case when people are involved, it is a good starting point for conceptual modeling of business processes. In an implementation, timing constraints can replace the greedy-runs-assumption. The timing constraints then not only define the time when a process performs its task, but also the maximum duration for which data may be buffered between the processes. Therefore such a channel behaves like a letter-box, as a buffer with a maximum time of delay.

## 2.2 Precise Definition of Business Processes

In the following, we will give a short introduction to the abstract syntax of business processes.

For simplicity, we assume that the type definitions for processes are given. Furthermore, for providing type definitions of ports, we assume a set of data sorts $S$ to be given as well. By $Pt$ we denote the set of all ports. With each port $pt \in Pt$, we can associate a data sort $Sort(pt)$ by
$$Sort : Pt \to S.$$

Above we argued that for reasons of methodical usage, a sort cannot always be assigned to every new port right away. Therefore, we allow for underspecification of ports, where sorts can be omitted at first, but may be associated with a port when appropriate.

A *process* is a tuple $(In, Out, b)$, where

- $In$ and $Out$ are disjoint sets of input and output ports and
- $b$ is the behavior relation describing the transformation from input to output. $b$ is defined as
$$b \subseteq \otimes_{p \in In} Sort(p) \ \times \ \otimes_{p \in Out} Sort(p)$$
where $\otimes$ is a large cross product collecting values of input and output ports, respectively.

$b$ is a relation describing the possible outputs of a business process, depending on its inputs. Such a relation may be described using natural language, as well as logic formulae or an algorithmic implementation. However, it can also be derived from a glass-box decomposition of the business process as we will show below.

Let in the following $BP$ be the set of business processes. For a specific process $bp \in BP$, we denote its sets of input and output ports by $In^{bp}$ and $Out^{bp}$, respectively. Furthermore, we define $Pt^{bp} = In^{bp} \cup Out^{bp}$ to be the set of all ports of process $bp$. As already noted, we usually suppress the port names in the diagrammatic representation. In Figure 3 only the interface ports have port names given explicitly ($in_1^{rb}$, $out_1^{rb}$, and $out_2^{rb}$)

Processes communicate with each other by exchanging messages via input and output ports. The sets of its input ports and output ports build the interface ($In^{bp}, Out^{bp}$) of a process $bp$. For simplicity we assume that the ports are unique for each process:

$$bp_1 \neq bp_2 \Rightarrow (In^{bp_1} \cup Out^{bp_1}) \cap (In^{bp_2} \cup Out^{bp_2}) = \emptyset$$

So far, we have treated each process in an isolated way. To describe complex system behavior, it is necessary to compose several single processes into a process network. This interconnection of processes corresponds to data dependencies.

A *business process net* is a tuple $(P, C, I, O)$ consisting of

- a set of business processes $P \subseteq BP$,
- a set of channels $C \subseteq Out \times In$ connecting some output port of a process to an incoming port of a successor process,
- a set of destinations $I \subseteq In$, for incoming ports, and
- a set of sources $O \subseteq Out$ for outgoing ports.

Note that the set of input and output ports in the overall $In$ and $Out$ are derived from the set of business processes $BP$ according to $In = \bigcup_{bp \in P} In^{bp}$ and $Out = \bigcup_{bp \in P} Out^{bp}$.

While $C$ specifies process connections by internal channels, $I$ describes through which ports of some internal process of the net, the environment may connect to the process net by providing input. Correspondingly, $O$ describes which ports are available for the environment.

In the following, we denote the set of business process nets by $BPN$.

To ensure that a business process net is well formed, several constraints must hold. By $\pi_1$ and $\pi_2$, respectively, we denote projection of a channel from $C \subseteq Out \times In$ to the sending or receiving process.

1. An input port is either internal to the process net, or a destination for incoming channels from the environment.

$$\pi_2(C) \cap I = \emptyset$$

2. All internal input ports within a process net are connected to exactly one output port within the net:

$$\forall c_1, c_2 \in C : \pi_2(c_1) = \pi_2(c_2) \Rightarrow c1 = c2$$

As a consequence of constraints 1 and 2, each input port is either assigned to exactly one output port, or is an input port for the whole net. Please note that we do not demand a similar restriction for output ports. Thus we support the modeling of output ports that feed into several channels, even up to broadcasting (see Figure 4).

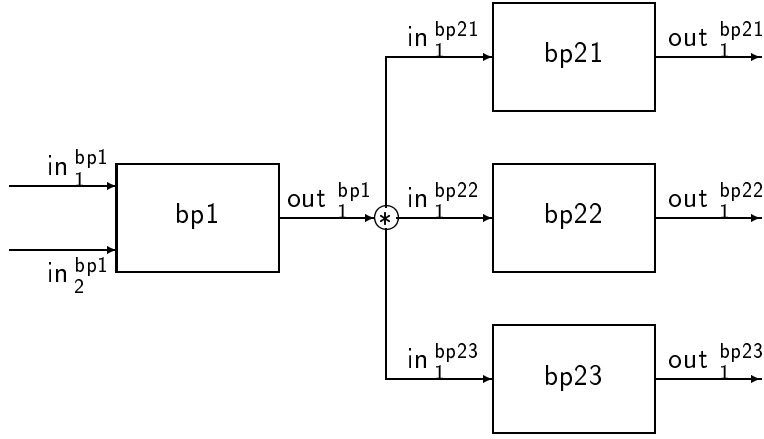

Figure 4: Broadcasting among processes

3. Port types of a channel fit together:
$$(pt_s, pt_d) \in C \;\Rightarrow\; Sort(pt_s) = Sort(pt_d)$$

4. We employ our process nets for modeling exemplary system behavior. Therefore, we want to exclude cyclic behavior, as this would add loops to the process model and thus contradict our intuition. Thus, we require the business process net to be acyclic, which we model by enforcing a possible serialization (quite like transactions can be serialized):
$$\exists f \colon P \stackrel{inj}{\to} I\!N : \forall bp_s, bp_d \in P : (Out^{bp_s} \times In^{bp_d}) \cap C \neq \emptyset \;\Rightarrow\; f(bp_s) < f(bp_d)$$

A business process net can be interpreted as a decomposition of a more abstract business process, providing a more detailed description of how the business process is realized. Thus, given a business process net $n = (P_n, C_n, I_n, O_n)$, we can easily build an abstraction of the decomposition description, yielding a black-box business process $bp \in BP$ with interface $In^{bp} = I_n$ and $Out^{bp} = O_n$, disregarding the internal connection structure.

Furthermore, the behavior of the composed business process $bp$ is derived from the behavior relation of the constituent business processes as well. This is done using a composition operation, quite similar to the functional composition known from mathematics. A more general composition operator which can also be used in this context is given in [PR97].

Please note that for a concrete representation of a business process net, e.g. when using an appropriate CASE-tool, we assume that it is possible to not visualize less important channels. Also, we assume that for several channels from one business process to the same destination process, a grouping into a single channel should be supported. This allows for a more compact process representation.

A complete *business process model* consists of a partial assignment of business process nets to processes that they refine. Thus, the set of business process models is defined by:
$$BPM = (BP \stackrel{par}{\to} BPN)$$

If we assume a top level business process called *system* to be given in the set of business processes $BP$, then a business process model consists of a hierarchy of business process nets. Here, some constraints do hold as well. The most obvious one is that the hierarchy of process nets has to be a tree, which furthermore needs to be finite in depth and breadth (not formalized here).

## 3 Refinement

In this section, we will introduce the notion of refinement. Then, we will give some transformation rules that are appropriate refinement steps for business process nets.

### 3.1 Notion of Refinement

A refinement step derives a new, more specific model from an existing one. *More specific* means that in all aspects, it provides *at least* the information that is provided by the original model. Therefore, all properties that can be deduced from the previous model are ensured to remain true.

From a technical point of view, we can classify refinement techniques in three different categories:

**Black-box to black-box refinement** allows to add information to an existing process definition, e.g. by extending its interface.

**Black-box to glass-box decomposition** (also called structural refinement) details a process definition by providing a corresponding process net, which describes the internal structure. This is the core of hierarchical decomposition of business processes. This kind of refinement strongly corresponds to the existence of a composition operation.

**Glass-box to glass-box refinement** modifies an already given process net into another one, e.g. by introducing additional dependencies between sub-processes.

Furthermore, we do not refine single processes or process nets in isolation, but the complete business process model. E.g. whenever a channel is added or changed within a process net, the sub-nets of the affected processes are also affected. The provided notion of refinement for business processes is related to behavioral and structural inheritance in the object-oriented context. For example, the ports of a business process, which is structurally decomposed, are inherited to sub-processes. Also, refinement through adding of dependencies (data-flow channels) is quite similar to the interface extension mechanism of inheritance.

Refinement can be achieved by applying it either independently, or in combination, to different parts of the business process model:

- the interface of business processes,
- the set of processes composed within a process net,
- the channel structure within a process net, and
- the data sorts.

This provides a large design space for different kinds of rules. The following set of rules by no means claims to be complete, but focusses on the most interesting ones. Practice will show that more rules are needed to refine business process models. However, it is very important to have rules that are semantically sound. A formal semantics for business processes was provided in [Thu98]. As we did not present a formal semantics for business process models in this paper, we will not formally prove the correctness of our rules. However, from the informal explanations of business processes it might become clear that the refinement rules are semantically sound.

As we have defined a business process model to be a hierarchy of business process nets, in general, refinement steps are transformations of the following kind

$$T : BPM \to BPM$$

Usually, the affected set of business processes is relatively small. It is restricted to those subtrees in the model hierarchy that comprise the glass box views of the modified processes. Quite often, only parts of those subtrees have to be actually modified. In the following, we will concentrate on the affected nets.

### 3.2 Refinement Rules

We start discussing the different possible refinement rules by regarding the business process in Figure 5.

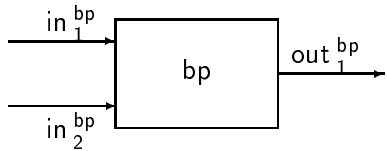

Figure 5: Simple business process

**Decomposition of a Business Process**

Assuming that process $bp$ is not yet decomposed by a sub-net in our business process model, then it is an obvious rule to allow the decomposition of $bp$ into a sub-net, as can be seen in Figure 6. If the result is a valid business process model (i.e. no forbidden loop occurs) then it is also a refinement of the former model.

This picture also clarifies why greedy components that evaluate partial information are useful. As sub-process $bp1$ does not rely on $in_2^{bp}$, it may greedily process its input before $in_2^{bp}$ exists. Similarly, sub-process $bp2$ may greedily start processing on input $in_2^{bp}$ without having to wait for process $bp1$ to produce its result.

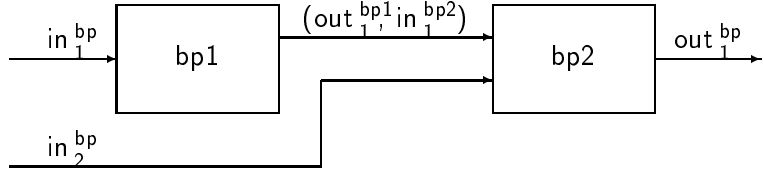

Figure 6: Decomposition of Figure 5 to process net with same interface

**Adding of Channels**

Adding of new channels is in general a valid refinement, as we assume that a business process model is typically rather abstract, depicting only the important, interesting dependencies or those that are well-known at a particular stage of modeling, and abstracting from unimportant dependencies. Strictly speaking, new channels are introduced by adding new ports to corresponding processes and connecting these ports via channels. Through refining a model into a more detailed one, these dependencies need to be explicitly added. Adding a channel always affects several layers of process nets. If the channel is internal to process $bp$, then it affects the glass-box description of $bp$ and the interfaces of its sub-processes (therefore black-box and glass-box on the sub-process level). The addition of a channel is allowed in either combination, it may even affect the interfaces of several layers, as long as no circular dependency relation is established, and process types are not contradicting. In Figure 7, a refinement of the net shown in Figure 6 and therefore a refinement of the original process is depicted.

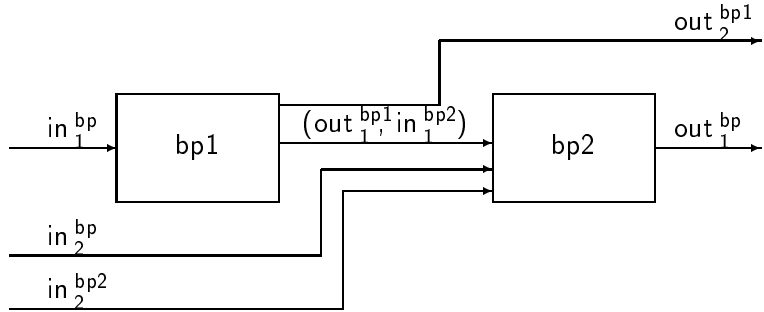

Figure 7: Refinement of Figure 6 through adding channels

**Decomposition of Dependencies, Data Refinement**

Dependencies are interpreted as dataflow channels, in which exactly one piece of data is transmitted. In general, this data document is quite complex and consists in itself of several different kinds of pieces of data. This may either be a record of relatively independent data or a collection (e.g. sequence, set) of data of the same type. As we originally allowed the data sort of a channel to be unspecified, it is a valid refinement to define the data sort of a channel. Figure 8 refines the process shown in Figure 5 by associating a complex data type with the process' output port.

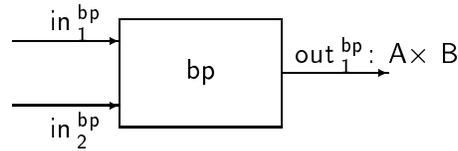

Figure 8: Refinement of Figure 5 through adding a data type

Besides specifying the data in greater detail, it would be even more useful to decompose a channel containing complex data into several channels with simpler data. This is e.g. useful after decomposing a process and realizing that different portions of the data channel are used in different sub-processes, or are collected from different sub-processes. Figure 9 illustrates a decomposition of process $bp2$ and output channel $out_1^{bp}$ (see Figure 6) into two new processes and new channels. Note that operator $\rightsquigarrow$ describes the relationship between the ports of the original process $bp$ to the corresponding ports of the refining subnet.

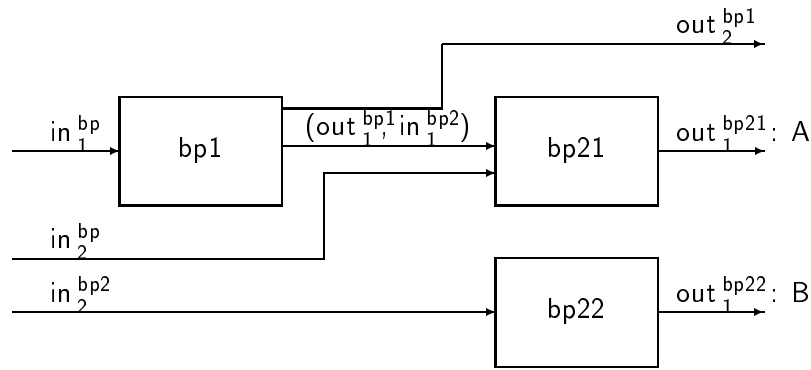

where $\quad out_1^{bp} \rightsquigarrow \{out_1^{bp21}, out_1^{bp22}\}$

Figure 9: Decomposition of an output channel from Figure 6

### Folding and Unfolding

Sometimes it becomes apparent that a given business process model is not well structured, as too many or too few dependencies between processes exist. This is frequently the case when new dependencies are added that the modeler has not been aware of previously. For restructuring process models, the basic transformation rules folding and unfolding are used. The unfolding rule basically copies a sub-net into a higher net, thus expanding it. Correspondingly, the folding rule does the inverse.

The combination of these two rules allows almost free rearrangements of business process models within a subtree, and therefore allows a rather flexible restructuring.

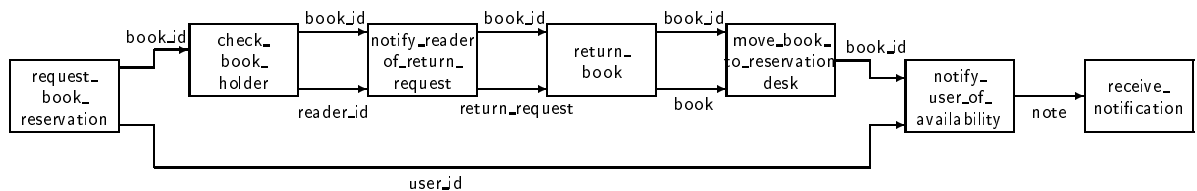

Figure 10: Unfolding of a process net

In Figure 10, such an unfolding is given, which expands the process net introduced in Figure 1 by the process net presented in Figure 3. In principle, unfolding of a process net is possible wherever a refining sub-net is defined for at least one of the processes in the process net.

While unfolding forgets structure, folding adds structural information. Therefore, some additional information, namely the set of folded processes, has to be provided in such a way that the result is still a well-formed business process model. Especially circular dependencies are not allowed.

Although folding and unfolding are in nature related with the composition rule, they deal with different methodical purposes. While composition is used to introduce new additional structure into a business process model, unfolding and folding deal with existing structure by flattening it or introducing more hierarchy, respectively.

Of course, combinations of these refinement rules lead to more powerful rules. However, it is also useful to provide specializations of the rules that e.g. deal with the addition of one channel, as this is often the case when tools are involved.

## 4 Outlook

In this paper, we have defined a formal notion of business processes and a set of rules that allow to refine them. We have demonstrated that it is possible to establish a precise notion of transformation, composition and refinement for business processes.

In the area of business process modeling there is a large amount of papers available that deal with different aspects of business processes [Ber98, Mil98, CKO92, Lew97, War94, HJ94, SL94]. However, business processes are often treated in an informal or at the most semi-formal way. Furthermore, a calculus of rules describing how to manipulate business processes in order to hierarchically refine them has not been defined so far.

The idea of refinement originally stems from algebraic specifications and was adapted to different forms of diagrams, e.g. to state machines in [Rum96] and to information flow architectures in [PR97].

To be more flexible, the notion of business processes in this paper can be enhanced further.

For example, when decomposing a business process, typically there is not a single one, but rather a set of different refining business process nets, from which a net will be chosen for execution according to some initial condition. These extensions can be simulated in our approach by sending simple signals (or null messages) indicating that no other information will arrive to those branches of the process net that shall not be executed, as well as by our notion of greedy evaluation for business processes.

However, a more explicit notion of selecting a process net for execution from a larger set of process variants would also allow for special refinement rules, e.g. the cutting of a computation path that will never be used. An integration of the refinement calculus presented here with the choice operators for input and output introduced in [Thu98] should form the basis for these more sophisticated refinement rules.

When implementing our notion of business processes, the greedy computation model is at least partly to be replaced by timing constraints. If, furthermore, explicit timing constraints are involved, then an analysis of critical paths may be of major interest.

Another very important aspect of implementing business process models is their mapping to components for execution. For executing a certain business process, the associated component adopts an appropriate role. The mapping of business processes to components is flexible, as one component can serve several business processes. Furthermore, it is also possible to assign the same role to several components, even if they belong to different types. On the other hand, a complex process may require the collaboration of several business components for its execution.

Consequently, the structure of components of the software system and the structure of the business process model may be to a large extent orthogonal. In addition to mapping business processes to components via roles, dataflow channels are to be considered as component internal or communication paths between components. A simple, yet often convincing solution is to constructively derive parts of the software architecture according to the business process model and to implement dataflow via shared database access and a separate control and scheduling mechanism.

Most important for a refinement calculus like the one given above is the preparation of appropriate tool support. Through implementation of a refinement calculus for state transition diagrams, as defined in [Rum96], we have shown the feasibility of such transformation tools [FR98].